# Zero-shifting Technique for Deep Neural Network Training on Resistive Cross-point Arrays

Hyungjun Kim[+], Malte Rasch, Tayfun Gokmen, Takashi Ando, Hiroyuki Miyazoe, Jae-Joon Kim[+], John Rozen and Seyoung Kim

IBM Thomas J. Watson Research Center, Yorktown Heights, NY 10598, USA

Email: sykim@us.ibm.com

[+]Current affiliation: Pohang University of Science and Technology, Pohang, Republic of Korea

## Introduction

In recent years, artificial intelligence (AI) has been melt into our everyday life in various forms. Among the several avenues in AI, deep neural network (DNN) has shown human level performance on several complex problems such as image classification [1,2] and speech recognition [3,4] due to increased computing power and Big Data [5]. However, training large DNNs with millions of parameters and large amount of data requires significant computing resources, often at the scale of data centers. Central Processing Unit (CPU) and Graphics Processing Unit (GPU) based conventional von Neumann architectures are widely used computing platforms for deep learning workloads as of today. Such compute units exhibit an intrinsic drawback, so-called "memory bottleneck", caused by the physical separation of memory and the compute engine. This requires continuous data transfer between two domains, which significantly limits the training performance. Therefore, novel architectures based on emerging memory technologies are recently proposed as alternatives.

A resistive memory device-based computing architecture is one of the promising platforms for energy-efficient DNN accelerators [6-24]. Fig. 1 shows the proposed concept using memory devices in cross-point array structure [6]. By supplying input values as voltage pulse on each row and setting weights between inputs and outputs as resistance at each cross-point device, one can perform fully-parallel vector-matrix multiplication in the analog domain where the results are the integrated current read at each column. This leads to significant acceleration in DNN inference where the vector-matrix multiplication is the most time-consuming operation [6]. However, using resistive crossbar arrays to perform the vector-matrix multiplication is a very old idea proposed more than 50 years ago [25] and it is not sufficient to accelerate DNN training workloads. Another critical step needed for DNN training is to use the device switching characteristics in order to perform the rank-1 update (outer product of two vectors) all in parallel with O(1) time complexity as well. As illustrated in [6], the rank-1 update of crossbar arrays can be performed by using stochastic pulses. Alternatively, deterministic pulsing schemes are also proposed to perform a similar rank-1 update on crossbar arrays [14].

The key technical challenge in realizing neural network training accelerator with resistive memory devices is to accumulate the gradient information in an unbiased way, and therefore, the device switching characteristics plays a critical role in determining the training performance of DNNs [6,18,19,21,27]. Unlike the digital numbers in software which can be assigned and accessed with desired accuracy, numbers stored in resistive memory devices in the form of resistance can only be manipulated following by the physics of the device. For example, resistive random access memory (RRAM) devices show asymmetric conductance update behavior which causes significant performance deterioration in DNN training [14,16,20,26]. Since the neural network algorithms were developed based on idealized systems that perform the necessary math using high accuracy digital numbers and unbiased compute units, any non-ideal device characteristics in hardware can impact the performance of the neural network. Therefore, additional techniques and algorithm-level remedies are required to achieve the best possible performance in resistive memory device-based DNN accelerators.

In this paper, we analyze asymmetric conductance modulation characteristics in RRAM by Soft-bound synapse model [28,29] and present an in-depth analysis on the relationship between device characteristics and DNN model accuracy using a 3-layer DNN trained on the MNIST dataset [30]. We show that the imbalance in up and down update caused by device asymmetry leads to poor network performance. We introduce a concept of symmetry point and propose a zero-shifting technique which is a method to compensate imbalance by programming the reference device and therefore changing the zero value point of the weight. By using this zero-shifting method, we show that network performance dramatically improves for imbalanced synapse devices.

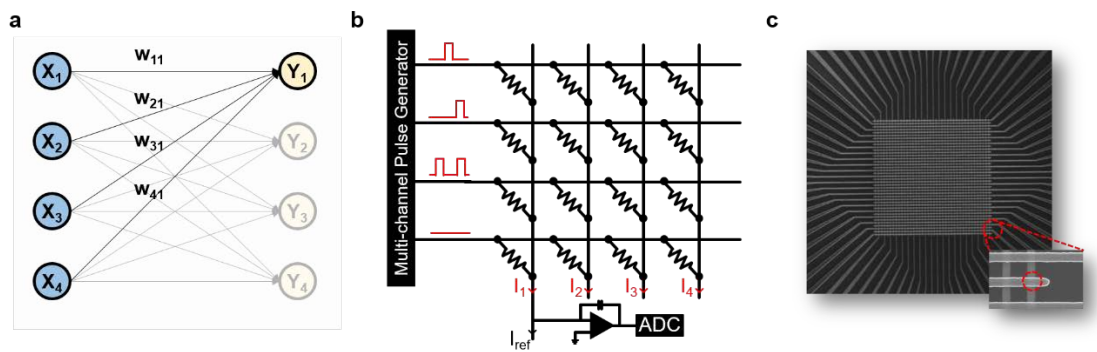

**Fig. 1** Resistive memory-based neural network accelerator concept with a synapse device array. **a** Schematic diagram of a fully-connected layer in a neural network with synaptic connections (W) between input (X) and output (Y). **b** RRAM cross-point array as synaptic weights. Inputs (X) of a layer are mapped to the number of voltage pulses in each row and outputs (Y) are represented by currents flowing through each column. **c** Photo of a fabricated 40x40 RRAM cross-point array.

**Analog Synapse model: Linear & Soft-Bound**

In this section, we elaborate on how we model analog synapse devices. When training a neural network in software using stochastic gradient descent (SGD), weights are updated iteratively, where each iteration linearly adds a calculated gradient to the weight. Thus, the amount of weight change per iteration is only dependent on the value of the gradient (times a learning rate), but not on the current value of the weight. This update behavior can only be implemented with resistive memory devices, when the synapse devices have a perfectly symmetric conductance response. However, most of the existing synapse devices instead show some form of nonlinear conductance responses. Fig. 2a shows an example conductance response of a RRAM synapse device. We applied 1000 successive pulses with positive voltage and 1000 successive pulses with negative voltage. Three cycles of potentiation and depression are shown in the Fig. 2a. It is obvious that both potentiation and depression are not linear and the conductance response depend on the current weight value. In particular, both potentiation and depression show saturating behavior when the magnitude of the weight value is large. This gradual saturation towards the bounds has been studied previously [28,29] and is called 'Soft-Bound behavior'. A Soft-Bound model can be expressed by following equations.

$$\begin{aligned} \text{Potentiation:} \quad & \frac{dw^+}{dn} = \frac{\Delta w_0^+}{w_{max}}(w_{max} - w) > 0 \\ \text{Depression:} \quad & \frac{dw^-}{dn} = \frac{\Delta w_0^-}{w_{min}}(w_{min} - w) < 0 \end{aligned} \quad (Eq.1)$$

Here, $\Delta w_0^+$ and $\Delta w_0^-$ represent the amount of weight change when the weight value is zero for potentiation and depression, respectively. $w_{max}$ and $w_{min}$ denote for maximum and minimum weights that a synapse device can represent (in a limit sense). We tried to fit the measurement data with the Soft-Bound model and the result is shown in Fig. 2b. Note that the Soft-Bound model fits the device data better than a linear model.

Fig. 2(c-f) show conductance responses of the linear model (c and d) and the Soft-Bound model (e and f), respectively. Fig. 2c and 2e show the resulting weight change as a function of the number of pulses given. Fig. 2d and 2f show the amount of weight change (delta weight) when one update pulse is given as a function of the current weight value. The linear model shows linear potentiation and depression, thus the amount of weight change is constant regardless of the current weight value (Fig. 2d). On the other hand, the conductance response in the Soft-Bound model depends on the current weight. The Soft-Bound model shows a linear relationship between current weight value and the amount of weight change per pulse (Fig. 2f). Therefore, the amount of conductance change for a single pulse is in general different for potentiation and depression and depends on the current weight value. When the device is in a high conductance state (high weight value), the synapse responds very weakly when potentiated even more, but responds with a large weight decrease when depressed. Similarly, when the conductance of the device is low, the synapse responses very weakly when depressed even more, but responds with large weight increases when potentiated.

Although the Soft-Bound model fits the device characteristics well, its gradual saturation behavior causes a weight dependent gradient update during SGD. To analyze the effect of such non-ideal conductance response during training, we simulated the neural network training with the Soft-Bound device model. As shown in Eq. 1, there are two parameters which determine the characteristics of the Soft-Bound model; $\Delta w_0^{\pm}$ and $w_{max/min}$. Since the ratio of $\Delta w_0^{\pm}$ and $w_{max/min}$ determines the slope of the conductance responses of a synapse device (Fig. 2f), identical device characteristics can be mapped to multiple combinations of $\Delta w_0^{\pm}$ and $w_{max/min}$. For example, a Soft-Bound synapse device can be used to represent a weight value between -1 and +1 with $\Delta w_0^{\pm}$ of 0.1. At the same time, the same physical device can be used to represent a weight value between -5 and +5 with $\Delta w_0^{\pm}$ of 0.5. We refer to these free parameter choices as different weight-to-devices mapping strategies. In the following section, we first show that depending on how the weight values are mapped to the devices, the network performance can vary dramatically, although the same physical device was used.

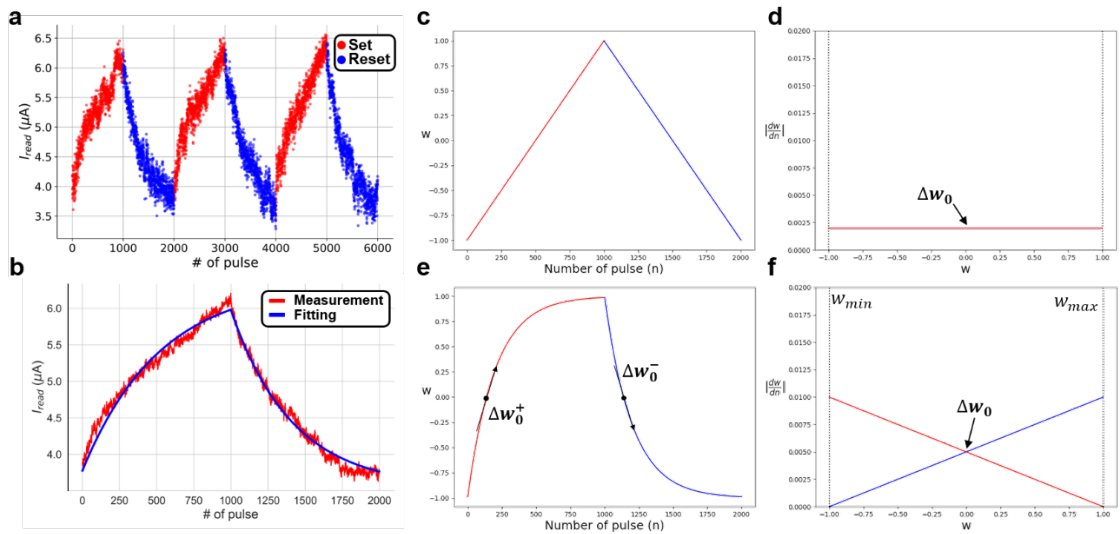

Fig. 2 **a** Conductance response during potentiation and depression in RRAM synapse devices (3 cycles). **b** Fitting experimentally obtained device data using the Soft-Bound model. Ten cycles of measurement data were averaged before fitting. Saturating characteristics are shown in both directions. **c** and **d** show conductance responses of a linear device model. Linear devices show linear conductance response as the number of write pulse increases. Since the model is linear, the amount of weight change is constant regardless of the current weight. **e** and **f** show conductance responses of the Soft-Bound model. The Soft-Bound model shows saturating behavior as shown in **e**. Unlike the linear model, the amount of weight change of the Soft-Bound model is not constant over the weight range. Instead, delta weight and current weight have linear relationship. As the absolute value of the slope increases, the conductance response of the device becomes more nonlinear.

**Training result using Soft-Bound device model**

      To evaluate the neural network performance depending on the Soft-Bound model parameters, we trained a 3-layer fully-connected network on the handwritten digit classification benchmark (MNIST [30]). The network has 784 input neurons and 256, 128, and 10 output neurons in the three layers, respectively. For activation function, we used a sigmoid function as in Eq. 2.

$$\sigma(x) = \frac{1}{1+e^{-x}} \quad (Eq.\ 2)$$

The network was trained for 30 epochs and the results were evaluated using 10,000 test images that the network has never seen during training. To simulate realistic device operation, we included physical non-idealities and variations into the training process. We used the same parameter setting as suggested in [6]. For example, we used 30% device-to-device (spatial) variation and cycle-to-cycle (temporal) variation of the $\Delta w_0^\pm$ parameter and 30% device-to-device (spatial) variation in the $w_{max/min}$ parameters.

      The training results using the Soft-Bound device model with device variations and analog noise are shown as contour plot in Fig. 3a. X and Y axes of the contour plot are $\Delta w_0$ and $w_{max}$ respectively. In this experiment, we assumed that $\Delta w_0^+$ and $\Delta w_0^-$ have the same absolute value $\Delta w_0$, and the absolute values of $w_{max}$ and $w_{min}$ are also identical while individual devices may show a 30% variation in both parameters as discussed above. We varied $\Delta w_0$ from 0.004 to 0.3 and $w_{max}$ from 0.5 to 20. Each data point in the contour plot represents the test error (%) after 30 epochs of training in log-color scale. As mentioned earlier, the ratio of $\Delta w_0^\pm$ and $w_{max/min}$ determines the shape of the Soft-Bound device model. Therefore, all the data points on a line (with thus constant ratio of both parameters) can be considered achievable by an identical physical device. For example, data points on the line marked by ⓑ400 are using the same device model shown in ⓑ of Fig. 2b. As the ratio $w_{max}/\Delta w_0$ increases, the Soft-Bound models' update behavior becomes more linear and indeed approaches the linear

model asymptotically. Thus, the region in the contour plot having large ratios result in very good network performance (blue region).

Meanwhile, we found that the training results might vary even if the same physical device model was used. In particular, the training results on a line with constant ratio (that are achievable with the same physical device) are typically not the same in the contour plot. For instance, in case of the ⓑ400 line, the test error rate is relatively high when both parameters are small. When both parameters increase, the test error becomes smaller, however, gets larger again as both parameters increase too much. In the first region where both parameters are small, the network performance is limited by the available range of weight values. Even though the device can be tuned very accurately with small $\Delta w$, the weights are restricted to have small values which limits the network performance. On the other hand, when both parameters are too large, the network performance is limited by inaccurate updates in the relevant weight range. Though weights can have wide range of values, single update of a weight results in abrupt change of weight, therefore accurate weight update is not possible. The best network performance can be achieved when the $\Delta w_0$ and $w_{max}$ have appropriate values. This observation gives us a lesson that one must make considerable effort on mapping network parameters on device characteristics to achieve better network performance.

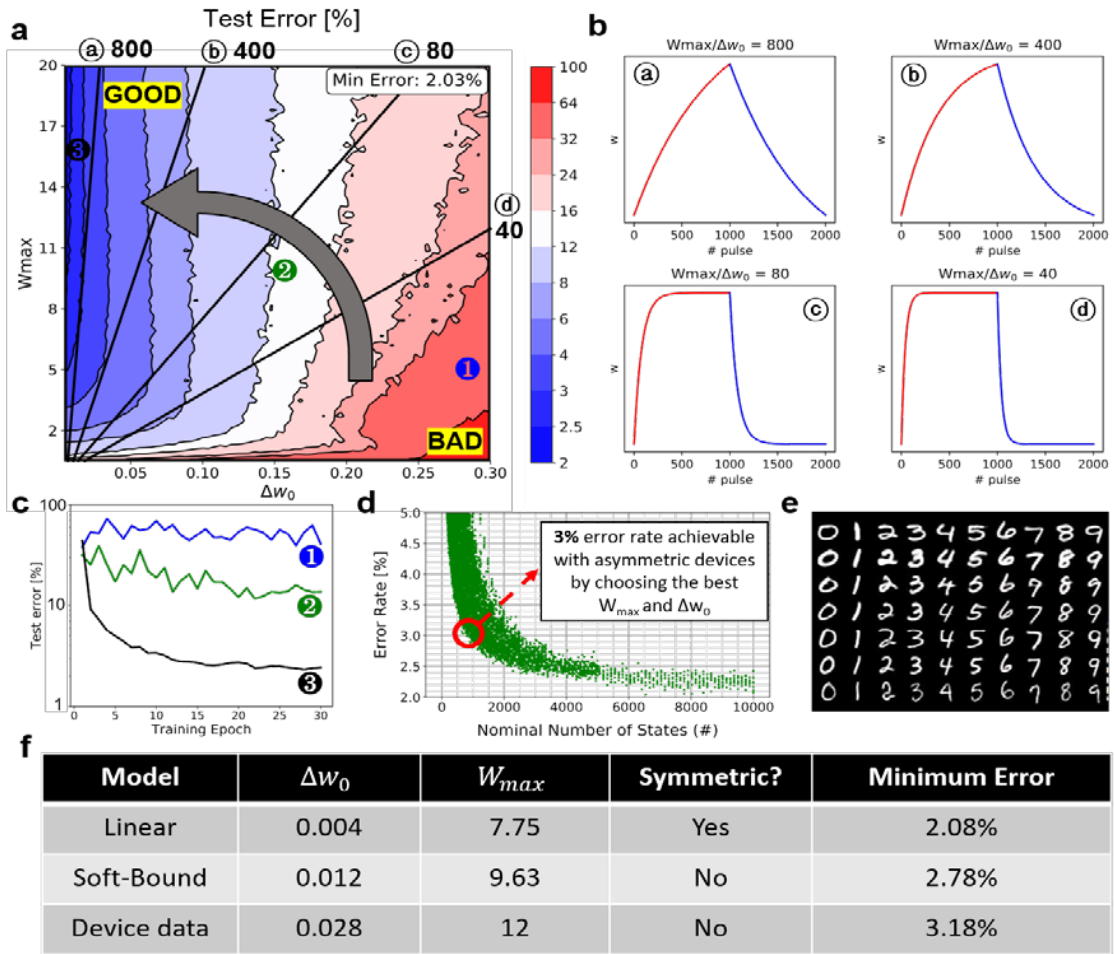

**Fig. 3 a** Contour plot of training result where $\Delta w_0$ and $w_{max}$ are varied. $\Delta w_0$ was swept from 0.004 to 0.3 and $w_{max}$ was varied between 0.5 and 20. Test error rate is depicted as color in log scale. Blue region represents low test error while red region represents high test error. **b** The ratio of $\Delta w_0$ and $w_{max}$ decides the shape of Soft-Bound model. When $w_{max}/\Delta w_0$ is large, the model is close to linear model. Data points on the same line are training results with identical device model but with different $\Delta w_0$ and $w_{max}$. **c** Training procedures for different data points in **a**. **d** Test error rate [%] as a function of nominal number of states. The nominal number of states is calculated following $(w_{max} - w_{min})/\Delta w_0$. As the model becomes linear, in other words as the nominal number of states increases, test error rate decreases. Even if the nominal number of states is fixed, the error rate may be different depending on the combination of $\Delta w_0$ and $w_{max}$. **e** Hand-written digit dataset (MNIST) used for the training. **f** Summary table for the training simulation. Even though the device model is not symmetric (Soft-Bound), same level of test error with the case using linear device model can be achieved. Furthermore, error rate of 3.18% can be achieved with actual device measurement data.

**Symmetry point in Soft-Bound model**

In this section, we introduce the notion of the symmetry point in the Soft-Bound model and its importance in neural network training. In the Soft-Bound model, there is one point where the graphs of potentiation and depression cross, as shown in Fig. 2d. We call the crossing point the symmetry point since the absolute values of $\Delta w_0^+$ and $\Delta w_0^-$ are the same in this point. If the Soft-Bound model has a symmetry point it is always unique. The reason why the symmetry point is important is that the conductance state of the Soft-Bound device tends to converge to the symmetry point with random programming pulses: When the current weight is smaller than the symmetry point, potentiation is stronger than depression. Therefore, the weight tends to increase towards the symmetry point. In contrast, when the current weight is larger than the symmetry point, depression is stronger than potentiation resulting in an overall decrease of the weight. Hence, if the device is updated randomly, the state of the device will converge to the symmetry point.

Having a hardware-related process that counter-acts the weight update and shifts the weight distribution towards a non-zero offset (the symmetry points) is obviously very detrimental in the training process. We here argue that if the introduced drift is towards zero instead, it would act like a typical regularization term and, therefore, likely less detrimental on performance. A drift toward zero can be achieved by mapping the weight range in accordance of the characteristics of the symmetry point. If the symmetry points were at 0, weights will tend to have zero values instead of a random offset. For most weight mapping strategies the Soft-Bound model will, however, be unbalanced, i.e. the symmetry point will not be at zero. In previous section, we assumed the special case that potentiation and depression are balanced which means that 1) the absolute values of $\Delta w_0^+$ and $\Delta w_0^-$ are same and 2) the logical zero weight is exactly in the middle of $w_{max}$ and $w_{min}$. However, some of the experimentally measured synapse devices do not follow such assumption as shown in Fig. 4. One of the potentiation and depression curves can be more nonlinear than the other curve as shown in Fig. 4. This behavior can be represented by different $\Delta w_0^+$ and $\Delta w_0^-$ values. In the case shown in the first graph of Fig. 4a, the absolute value of $\Delta w_0^-$ is larger than that of $\Delta w_0^+$ which indicates the update is more abrupt when down. When the model is balanced, the symmetry point is at w=0. However, if the model is not balanced as in Fig. 4a, the symmetry point is not at w=0.

To evaluate the effect of imbalance in the model systematically, we use the weight value of symmetry point ($w_{sym}$) to indicate how unbalanced the model is. Since we know $w_{max/min}$ and $\Delta w_0^{\pm}$, we can derive $w_{sym}$ using following equation.

$$w_{sym} = \frac{|\Delta w_0^+| - |\Delta w_0^-|}{\frac{|\Delta w_0^+|}{w_{max}} - \frac{|\Delta w_0^-|}{w_{min}}} \quad (Eq. 3)$$

$w_{sym}$ indicates how unbalanced the model is since it simply represents how far the symmetry point is from logical zero weight. When $w_{sym}$ has a negative value, the potentiation of the device shows more linear characteristics than the depression. On the other hand, when $w_{sym}$ has a positive value, the potentiation of the device shows more nonlinear characteristics than the depression. Note that the balancedness of a model is different from the notion of the symmetry of a model. Soft-Bound synapse models are always asymmetric, however, they can either be balanced or unbalanced.

With different $w_{sym}$ values, we analyzed the effect of update imbalance in similar way as in the previous section. In this case, $\Delta w_0$ denotes the mean value of $|\Delta w_0^+|$ and $|\Delta w_0^-|$ since $|\Delta w_0^+|$ and $|\Delta w_0^-|$ may have different values. Fig. 4c shows the contour plots with different $w_{sym}$ values. The plot in the center with $w_{sym}$=0 is identical with Fig. 3a. As the model becomes unbalanced, the overall training performance dramatically deteriorates (red region broadens). In other words, the network performance gets degraded as the synapse device becomes more unbalanced. Note that, even if the Soft-Bound model parameter would be mapped to the devices by globally adjusting the average across devices to be balanced, device-to-device variations in the parameters would nevertheless cause individual devices to be unbalanced.

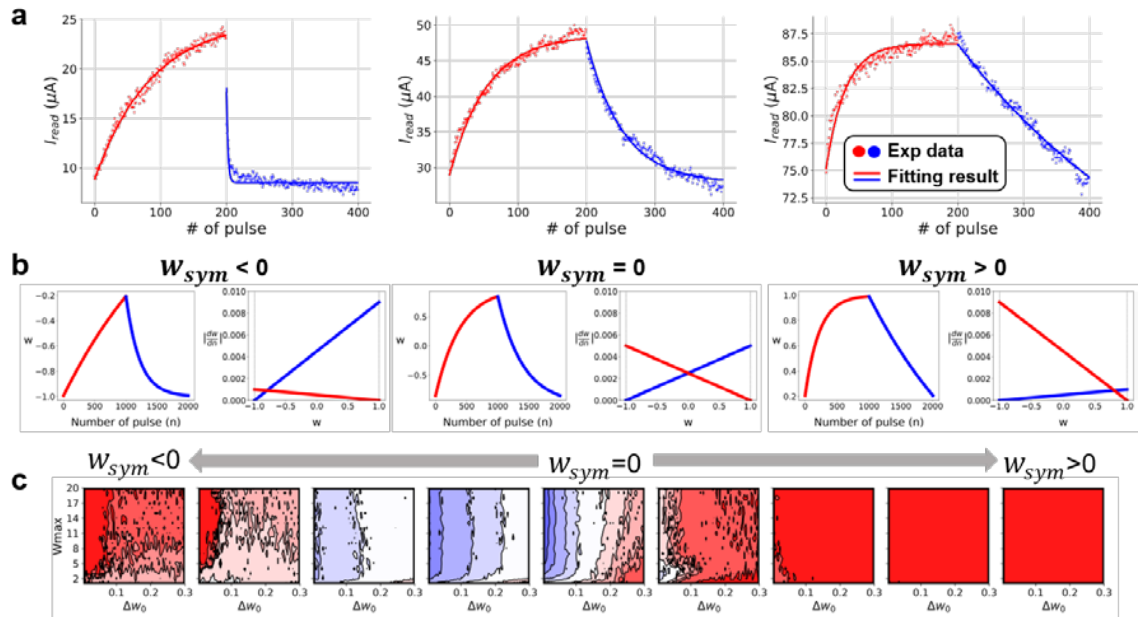

**Fig. 4** Unbalanced Soft-Bound models and its effect on network performance. The degree of nonlinearity for potentiation and depression can be different for devices as shown in **a**. The device shown in the left has nonlinear potentiation and relatively linear depression. On the other hand, the device shown in the right has nonlinear depression and relatively linear potentiation. **b** Three different cases with different $w_{sym}$ are shown. Conductance response and its derivative are shown in a set. The intersection in derivative plot is called symmetry point since the amounts of delta weight for both potentiation and depression are same in such condition. When $w_{sym}$ is positive, potentiation is more nonlinear than depression. On the other hand, when $w_{sym}$ is negative, depression is more nonlinear than potentiation. **c** The contour plots of training results with different $w_{sym}$ values are shown. As $w_{sym}$ moves away from zero, the overall test error increases.

**Zero-shifting technique to compensate imbalance**

In this section, we propose a technique called 'zero-shifting' to compensate the imbalance of the Soft-Bound model on a physical device level. When a device model has non-zero $w_{sym}$ value, the weight distribution tends to shift toward the non-zero symmetry point. We saw in the last section that the shifted distribution of the weight typically causes significant degradation in network performance; therefore, we propose a compensation technique to achieve better performance even with poor device characteristics.

Since the main reason why the network shows poor performance is that the symmetry point is not at w=0 point, the proposed zero-shifting method is to shift the weight range to match the w=0 point and the symmetry point of the device. We keep the scale of the weight range and only shift the range as shown in Fig. 5. Re-mapping of the parameters is required for the zero-shifting process and described in the following equations.

$$\begin{aligned} w'_{min} &= w_{min} - w_{sym} \\ w'_{max} &= w_{max} - w_{sym} \end{aligned} \quad (Eq.4)$$

After zero-shifting technique is applied, $w'_{max}$ and $w'_{min}$ do not have the same absolute value anymore. Fig. 5 shows the device parameter mapping before and after zero-shifting technique. After zero-shifting technique is applied, $w_{sym}$ becomes 0. Since we shifted the weight range, negative and positive weight bounds may have different values.

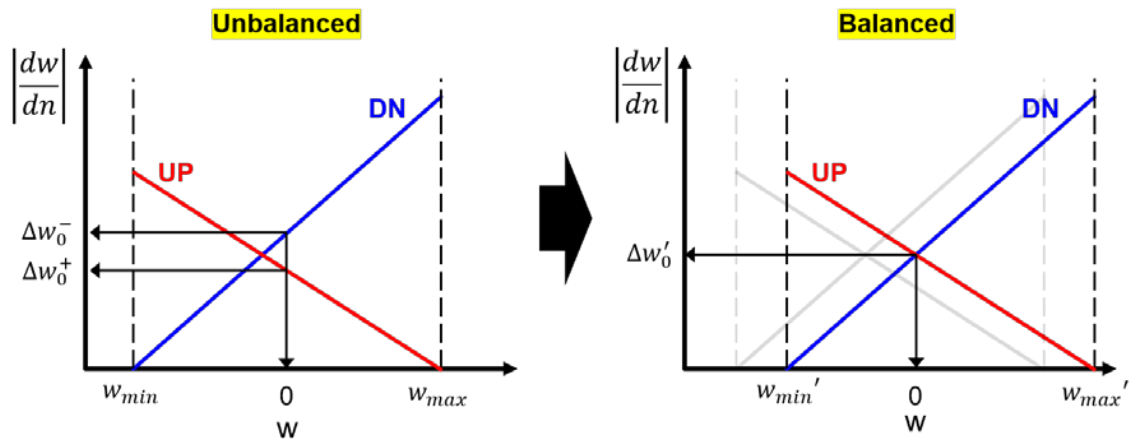

**Fig. 5** Proposed zero-shifting method to compensate the imbalance of the model. As shown in the left graph, the symmetry point of an unbalanced model is not at w=0. The proposed method re-maps the w=0 point to the symmetry point of the unbalanced model so that $w_{sym}$ becomes zero and the model balanced after compensation.

We further propose architecture and methodology to implement the proposed zero-shifting technique on a hardware level (Fig. 6). To map the weight zero to the symmetry point for each device, we propose to use two synapse devices per weight. Thus, there exist two separate cross-point arrays with identical size. One is to store the weight values, that are updated during training, and the other is fixed and for reference. Reference devices will be programmed once at the beginning to represent the symmetry point of their paired device. During the training, results from each array are subtracted from each other to derive the final output of the vector-matrix multiplication. When the proposed zero-shifting technique is applied, devices in the reference array must be programmed to the conductance that corresponds to the shifted zero weight. The detailed process is explained in Fig. 6. First, the devices in the weight array are updated until the weight of each device converges to its symmetry point. Unit pulses for potentiation and depression are given to all the devices repeatedly. After convergence, each device will be close to their own symmetry point. Then, the resulting conductances are copied to the reference array. In this way, conductance values of the reference array become the symmetry point of each paired device. Since this technique is adaptive, it is applicable without knowing $w_{sym}$ values of each device. In addition, the device-to-device spatial variation of the model imbalance can also be compensated.

Fig. 7 shows the detailed procedure of finding the symmetry point. Two exemplary cases with $w_{sym}<0$ and $w_{sym}>0$ are shown in Fig. 7a. Even though $w_{sym}$ and the initial state of each device are different, their conductance states converge to the symmetry point after enough potentiation and depression. We also demonstrate experimental result of finding the symmetry point with a fabricated RRAM device. Fig. 7b shows the conductance change during potentiation, depression, and the process of finding the symmetry point. Conductance of the device converges to a certain state where the size of the average weight change for potentiation and depression are the same.

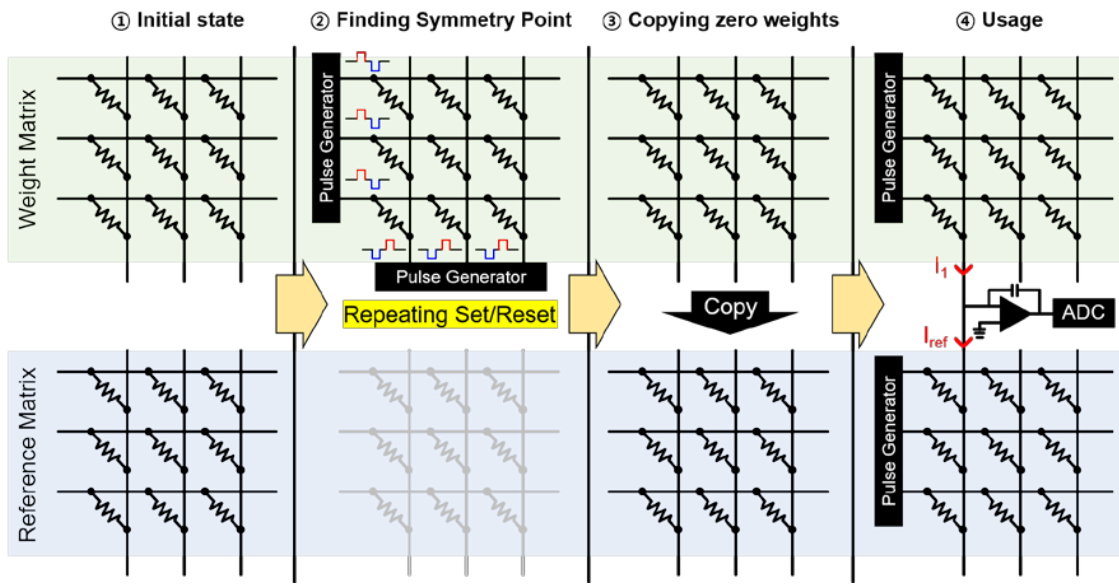

**Fig. 6** Schematic description of hardware implementation of the proposed zero-shifting method. Two crossbar arrays are required to represent positive and negative weights. A reference matrix is used to generate reference current with zero weights. Since each device in the weight matrix may have a different symmetry point, in other words, conductance of the device that represents zero weight may be different, the first step of the proposed method is to find the symmetry point of each device of the weight matrix. Positive and negative voltage pulses are applied to all the rows and columns of the cross-point array iteratively. After sufficient number of applied pulses, the conductance of each device will be close to the symmetry point of each synapse device. Next step is to copy the resulting conductance states of synapse devices into the corresponding devices of the reference matrix. By subtracting each device with its corresponding reference device, the effective symmetry point of each device in weight matrix is close to zero, therefore zero-shifting technique can be realized.

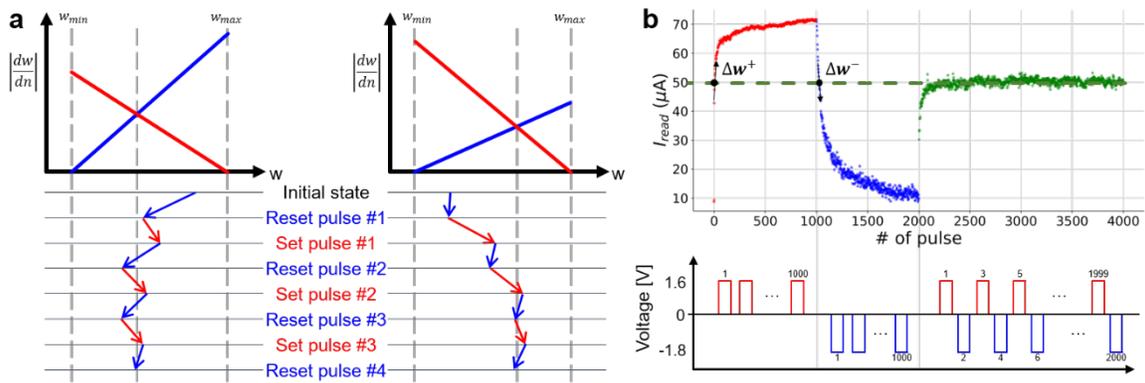

**Fig. 7 a** Procedure of finding the symmetry point with iterative set/reset pulses. Two exemplary cases of finding the symmetry point of each device are demonstrated. Note that both $w_{sym}$ and the initial states are different in the two cases. As the device experiences set and reset pulses iteratively, the conductance states(i.e.weights) of the device tends to converge to the symmetry point. **b** Experimental data of potentiation, depression, and finding the symmetry point. When the device experiences set/reset pulses repeatedly, its state converges to the symmetry point where delta weights are equal for potentiation and depression. Corresponding voltage pulses are depicted below the graph.

Fig. 8 shows the training results when the model imbalance is compensated by the zero-shifting technique. The overall network performance has been greatly improved even if the model becomes very unbalanced as shown in Fig. 8a. Minimal test error that can be achieved within the given domain are plotted as a function of $w_{sym}$ in Fig. 8b. When the model is unbalanced, the minimum error dramatically increases as $w_{sym}$ moves out of zero. However, if the zero-shifting technique is applied, the minimum error are almost constant regardless of the $w_{sym}$ value. Fig. 8c and d show the weight distributions of the last layer of network after training is done. When the model is unbalanced and $w_{sym}$ is not zero, the mean of the distribution is not at zero anymore (Fig. 8c). In contrast, when zero-shifting technique is applied, the mean of the distribution is still at zero even though the model is highly unbalanced (Fig. 8d). This observation makes it clear that the zero-shifting technique helps training the network by calibrating the weight distribution during the training process considering the imbalance in Soft-Bound synapse model.

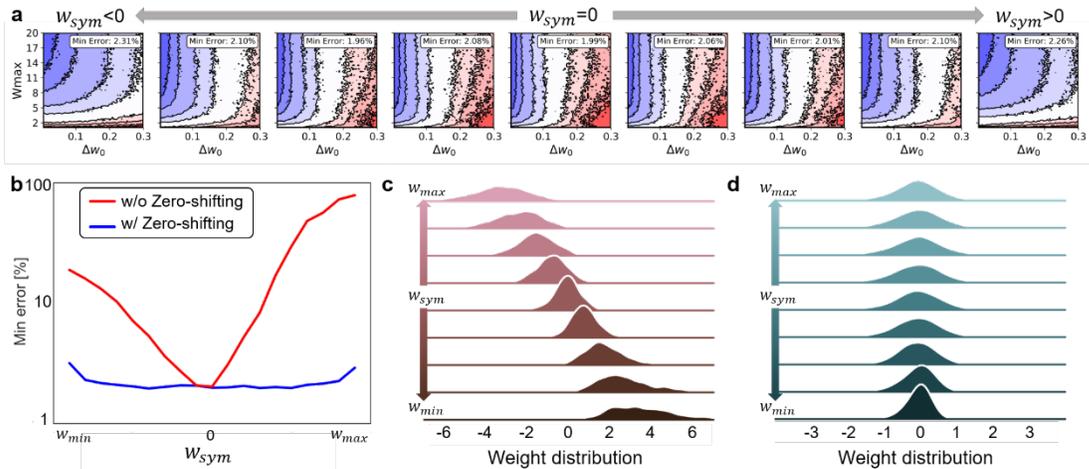

**Fig. 8** Training results with the proposed zero-shifting technique. **a** The contour plots of the training results with different $w_{sym}$ values. Imbalance of the model was compensated so that weight equals to zero at symmetry point for all cases. Compared to the results in Fig. 4, overall training results are much better when the proposed method is applied. Even if the model is highly asymmetric, minimum error rate of <3% can be achieved. **b** Minimum error rate for each $w_{sym}$

value before and after the zero-shifting method is applied. Unlike the unbalanced case, minimum error rate is greatly reduced especially when the model is unbalanced. **c** Weight distribution of the last layer after 30 epochs of training when zero-shifting method is not applied. Since the symmetry point is not at w=0 when $w_{sym}\neq0$, mean value of the weights is also not zero. **d** On the other hand, when the imbalance is compensated by the zero-shifting technique, the weight distribution is centered at w=0 regardless of the initial $w_{sym}$.

**Conclusion**

To date, many research groups have studied resistive memory-based neural network accelerators. However, research has been focused on improving the synaptic devices material properties to come closer to have ideal characteristics such as linear and symmetric conductance responses. Few works have reported that the non-ideal conductance response of RRAM devices may result network performance degradation during training, but they were focusing on relatively limited cases [6,14,26]. While these works were ceased concluding that it is important to develop synaptic devices with ideal characteristics for better network training, we further analyzed how these non-ideal characteristics (i.e. imbalance) affect network performance by extensive simulation on various parameter domains.

We first found that the network performance may vary depending on how to map the device parameters to network parameters, even if identical devices are used. We analyzed the effect of 1) maximum/minimum weight range ($w_{max/min}$) and 2) minimum weight change ($\Delta w_0^\pm$) on network performance. Even if the range of weight is wide enough, too large $\Delta w_0^\pm$ degrades the network performance since accurate weight update is not possible. On the other hand, even if $\Delta w_0^\pm$ is small enough for accurate weight update, a too small weight range also degrades the network performance. We found that there exists optimal values for the weight range and for the minimum weight change and finding this optimal point is essential to achieve good network performance.

We also proposed a zero-shifting technique that can compensate the imbalance of the potentiation and depression. Synapse devices may show different characteristics for potentiation and depression throughout its conductance range. When potentiation (or depression) is much stronger than depression (or potentiation), $w_{sym}$ has positive value (or negative value) and the network performance becomes very poor. The proposed zero-shifting technique shifts the weight range so that the symmetry point of the device matches with the conductance state where weight equals to zero. During neural network training, the conductance state of Soft-Bound synapse model tends to converge to symmetry point as the weight is being updated. The proposed zero-shifting technique utilize the effect of non-ideal conductance response to improve the network performance.

Finally, there are several interesting future directions. First, it would be very helpful to consider different parameter mapping for each layer in a network. In general, weight range of a layer has strong relation with the size of the layer. Each layer might have different optimal weight range and its effects on the network performance might be different [19]. Second, it is worth trying to analyze other device models with similar methodology. Not all the devices show Soft-Bound behavior and other device models may show different results. Third, evaluation on larger network and dataset is another interesting extention. Convolutional neural network (CNN) may be more sensitive to the non-linear conductance responses. Lastly, hardware-level implementation and demonstration of the proposed technique will be a very important milestone for RRAM-based neural network accelerator design.

**Methods**

We fabricated 2-terminal oxide RRAM with device dimensions from 150 x 150 µm$^2$ down to 200 x 200 nm$^2$. First, a SiO$_2$ underlayer was grown on a 200 mm Si wafer. Then, a 100 nm-thick TiN film was deposited by reactive sputtering as a bottom electrode and patterned, followed by deposition of a HfO2 layer by atomic layer deposition as a switching layer where a current conducting filament is formed. Next, a 20 nm-thick TiN was deposited by reactive sputtering as a top electrode. The device area was defined by photolithography and reactive ion etching of the TiN electrode. To perform zero-shifting technique on our ReRAM devices, we applied a sequence of weight update (write) pulses with the same voltage amplitude for each polarity. We used high-resolution source measure unit (SMU) to read the device conductance state between the write pulses. The write pulses had duration of 5 ns (unless otherwise mentioned) and various voltage amplitudes (typically for set pulse: 1.5–2.5 V; reset pulse: −2.0 to −3.0 V) were compared to investigate the impacts on switching characteristics.

**Simulation of neural networks**

We used custom framework based on Caffe2 [31] for RRAM-based neural network training. Three layer fully-connected network was used for classifying the hand-written digit (0~9) dataset (MNIST). During training phase, all 60,000 images in the training set were shuffled and fed to the network. We trained networks for 30 epochs, therefore each image was shown 30 times during training. At the end of every training epoch, 10,000 images in the test set were shown to the network and the classification accuracy was evaluated. An image in the MNIST dataset has 28x28 pixels, where each pixel has grayscale value from 0 to 255. There are 784 (=28x28) input neurons, 256 and 128 hidden neurons in the hidden layers, and 10 output neurons. The first and second fully-connected layers are followed by activation functions which can be sigmoid or hyperbolic tangent. To compute the score of each output neuron, we used a softmax classifier based on cross-entropy loss function. We did not use other layers such as batch-normalization or dropout. For parameter optimization, we used stochastic gradient descent (SGD) with mini-batch of 1. The initial learning rate was 0.01 and scaled by 0.5 every 10$^{th}$ epoch. Since stochastic behaviors of RRAM devices act as strong regularizer, we did not use weight decay during training. We also considered the effect of circuit-level noises. We assumed 9-bit Analog-to-Digital Converters (ADCs) and 5-bit Digital-to-Analog Converters (DACs). In addition, we added Gaussian noise with standard deviation of 0.06 to the analog weighted-sum value.

## Supplementary material
**Effect of asymmetric activation function**

Another interesting observation from Fig. 4 in main script is that the effect of imbalance in the device model is not symmetric (Fig.4c). When $w_{sym}$<0, there still exists a blue region where relatively good training results can be achieved while the training results for entire domain are poor when $w_{sym}$>0. We found that this phenomenon is due to the asymmetric activation function used in the network of the main text. As mentioned earlier, we used sigmoid function as activation function in each layer. Sigmoid function is a nonlinear function in which the output is always positive value as shown in Fig. S1a. Since the activation values are always positive, weights had better have negative values, otherwise the outputs of a layer will be always positive. When a device model has negative $w_{sym}$ value, the weight tends to have negative value rather than positive value. On the other hand, when a device model has positive $w_{sym}$ value, the weight will also have positive value as activation value; therefore capability of the network will be degraded. To prove that the asymmetric effect of imbalance is due to the asymmetric activation function, we trained the same network with symmetric hyperbolic tangent (Tanh) activation function. Tanh function is similar to sigmoid function but show symmetric characteristics over the origin point as shown in Fig. S1a. The training results are shown in Fig. S1b. As expected, the effect of the model imbalance is more symmetric than in the case of using the sigmoid function.

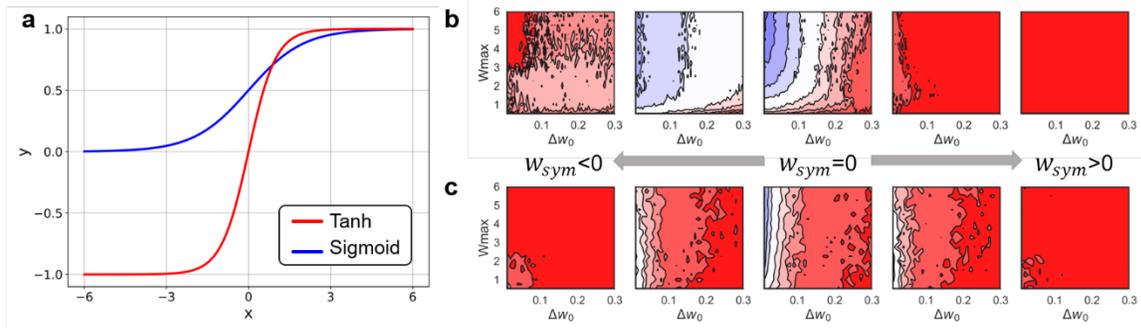

**Fig. S1 a** The two different activation functions compared for this simulation. The Sigmoid function is not symmetric around the origin while Tanh function is symmetric around the origin. **b** Effect of the symmetry point when asymmetric activation function (sigmoid) is used. The effect of $w_{sym}$ is not symmetric for $w_{sym}>0$ and $w_{sym}<0$. **c** Effect of the symmetry point when a symmetric activation function (Tanh) is used. In this case, the effect of $w_{sym}$ is also symmetric.